\documentclass[a4paper]{jpconf}
\usepackage{iopams}
\usepackage[comma,numbers,square,sort&compress]{natbib}

\usepackage{graphicx}
\usepackage{siunitx}

\newcommand{\dd}{\ensuremath{\text{d}}}
\newcommand{\ddel}[1]{\ensuremath{\partial_{#1}}}

\newcommand{\hxy}{\ensuremath{h_{xy}}}
\newcommand{\hxyh}{\ensuremath{\hat{h}_{xy}}}
\newcommand{\Qv}{\ensuremath{Q\!\left(v\right)}}
\newcommand{\Qsqv}{\ensuremath{Q^2\!\left(v\right)}}

\newcommand{\mss}{\ensuremath{m_\text{s}}}
\newcommand{\ts}{\ensuremath{t_\text{s}}}
\newcommand{\sigs}{\ensuremath{\sigma_\text{s}}}

\newcommand{\Ap}{\ensuremath{A_\text{p}}}
\newcommand{\tp}{\ensuremath{t_\text{p}}}
\newcommand{\sigpv}{\ensuremath{\sigma_\text{pv}}}

\begin{document}
\title{AdS/CFT far from equilibrium in a Vaidya setup}

\author{Michael F Wondrak$^{1,2}$, Matthias Kaminski$^{3}$, Piero Nicolini$^{1,2}$ and Marcus Bleicher$^{1,2}$}

\address{$^1$ Frankfurt Institute for Advanced Studies (FIAS), %
\\Ruth-Moufang-Str.~1, 60438 Frankfurt am Main, Germany}
\address{$^2$ Institut f\"ur Theoretische Physik, Johann Wolfgang Goethe-Universit\"at Frankfurt am Main, %
\\Max-von-Laue-Str.~1, 60438 Frankfurt am Main, Germany}
\address{$^3$ Department of Physics and Astronomy, University of Alabama, %
\\514 University Blvd., Tuscaloosa, AL 35487-0324, USA}

\ead{wondrak@fias.uni-frankfurt.de, mski@ua.edu, nicolini@fias.uni-frankfurt.de, bleicher@fias.uni-frankfurt.de}

\begin{abstract}
In this paper we apply the AdS/CFT correspondence to study a strongly coupled plasma far from equilibrium with a strong emphasis on the shear behavior. The plasma serves as a model for an electrically charged quark-gluon plasma. On the gravitational side, we use an ingoing Vaidya black brane spacetime. The highest rate of mass infall is confined to a short time interval.
\end{abstract}

\section{Introduction}
\label{sec:Introduction}
According to present cosmological models, the universe was radiation-dominated after the inflation phase and while expanding it cooled down following the law $T \propto t^{-1/2}$ where $T$ is the temperature and $t$ is the time after the big bang. Until about $\SI{e-5}{\s}$ the temperature exceeded a value of $\SI{200}{\MeV}$ and the universe was filled with a very special state of matter called the quark-gluon plasma (QGP) \cite{Mukhanov2012}. Such a state is supposed to be present in the interior of neutron stars and in supernova explosions. The QGP describes highly compressed baryonic matter at high temperatures subject to quantum chromodynamics (QCD). QCD has a large coupling constant at low energies which implies that the fundamental degrees of freedom, quarks and gluons, do not manifest individually, but are confined to hadrons. At larger energies corresponding to temperatures close to the critical temperature $T_\text{c} \sim \SI{200}{\MeV}$, the coupling decreases so that quarks and gluons deconfine and form the QGP. Finally, at even higher energies, the theory becomes weakly coupled.

The treatment of the QGP around $T_\text{c}$ is challenging because the QCD coupling constant is too large to allow for perturbation theory. There is much activity to describe the QGP by, \textit{e.g.}, hydrodynamics, kinetic theory, effective field theory, and lattice QCD (for a detailed treatment cf.~\textit{e.g.}~\cite{CBMPhysicsBook2011}). A particular approach is provided by the AdS/CFT correspondence. It connects a supergravity theory on an asymptotically Anti-de Sitter (AdS) spacetime with a non-abelian conformal field theory (CFT) on the flat conformal boundary of that spacetime. As a strong-weak duality, the gravitational spacetime is weakly curved, while the conformal field theory is strongly coupled.
Even more general versions have been conjectured, the strong and strongest form, which also allow stronger curvature on the gravity side and smaller couplings on the field theory side. Furthermore, the gauge/gravity duality can also address non-conformal fields on the boundary (cf.~\textit{e.g.}~\cite{AmmonE2015}).

On the experimental side, a QGP has already been produced in heavy-ion collisions at the Relativistic Heavy Ion Collider (RHIC) and the Large Hadron Collider (LHC) which was proven by observation of elliptic flow and jet quenching \cite{GyulassyM2005,ATLAS2010}.
Experimental data allows to extract how the highly inhomogeneous initial state after the collision comes to local equilibrium and how it reacts to fluctuations. 
The corresponding transport coefficients serve as benchmarks for theoretical approaches towards the QGP. One prominent representative is the ratio of shear viscosity to entropy density, $\eta/s$. Its extrapolated value lies close to the AdS/CFT prediction of $1/(4\pi)$ for gauge theories at large 't Hooft coupling which possess a supergavity dual \cite{KovtunSS2003,KovtunSS2005,BuchelL2004}.

While the hydrodynamic description requires local thermodynamic equilibrium and is therefore applicable only at advanced times after the heavy-ion collision, the AdS/CFT correspondence can be used at earlier times in the far-from-equilibrium regime \cite{CheslerY2008,CheslerY2009,BalasubramanianBCKKMTV2013,DavidK2015,%
BanerjeeIJMR2016,Ishii2016}.

We focus our interest on determining the shear behavior of a hot, charged, strongly coupled plasma far from equilibrium. 
The paper is organized as follows: In section \ref{sec:AdS_CFTCorrespondenceAndHydrodynamics} we introduce the quantities relevant for our calculation and show relations among them based on the AdS/CFT correspondence. Moreover, we discuss shear for long-scaled perturbations in the hydrodynamic regime. A sketch of our calculation procedure together with the spacetime model can be found in section \ref{sec:CalculationProcedure}. Section \ref{sec:MetricFluctuation} focuses on the evolution of fluctuations using numerical methods. We conclude in section \ref{sec:Conclusion}.

\section{AdS/CFT correspondence and hydrodynamics}
\label{sec:AdS_CFTCorrespondenceAndHydrodynamics}
The AdS/CFT correspondence allows to investigate a quantum field theory at finite temperature by studying a black brane, \textit{i.e.}, an infinitely extended black hole with planar topology in a higher-dimensional product spacetime which consists of an asymptotically AdS spacetime and a compact manifold \cite{Witten1998}. The field can be thought to be located on the $d$-dimensional boundary of the ($d$+$1$)-dimensional AdS bulk spacetime. Since it is possible to associate thermodynamic properties to black objects (black hole thermodynamics), several quantities can be compared directly.
For example, the temperature of the field on the boundary, $T$, is given by the Hawking temperature of the black brane, $T_\text{H}$. The entropy of the field state equals the Bekenstein-Hawking entropy, $S_\text{BH}$. Therefore the entropy density $s$ follows from the ratio between $S_\text{BH}$ and the field extension, which is $\text{Vol}\!\left(\mathbb{R}^{d-1}\right)$ for the Poincar\'{e} patch:
\begin{equation}
T = T_\text{H} = \frac{\kappa}{2\pi},
\qquad\qquad
s = \frac{S_\text{BH}}{\text{Vol}\!\left(\mathbb{R}^{d-1}\right)}
=\frac{A}{4G\,\text{Vol}\!\left(\mathbb{R}^{d-1}\right)}.
\end{equation}
Here $\kappa$ and $A$ denote the surface gravity and the horizon area of the black brane and $G$ is the higher-dimensional gravitational constant.

The AdS/CFT correspondence or, more generally, the gauge/gravity duality states the equality of the grand-canonical partition function of the field theory on the boundary at $r\to\infty$ and that of supergravity in the bulk. The GKP-Witten relation dynamically connects the field quantities, \textit{e.g.}, a U($1$) current on the boundary with a U($1$) bulk field. 
By choosing a non-vanishing electric chemical potential $\mu$ on the boundary, the field state becomes electrically charged. The electrodynamic U($1$) gauge symmetry on the boundary is modeled by the trace of the maximal abelian subgroup of the SU($\mathcal{N}$) R-symmetry where $\mathcal{N}$ is the number of supercharges present \cite{AmmonE2015}. The bulk gauge field $A_m$ gives rise to the chemical potential by its boundary asymptotic behavior, 
\begin{equation}
\mu = {\left.A_0\right|}_{r_\text{h}}-{\left.A_0\right|}_{r\to\infty}
\end{equation}
where $r_\text{h}$ denotes the position of the black brane apparent horizon. 
The energy-momentum tensor of the field theory is sourced by the boundary value of the bulk metric $g_{mn}$. Since shear introduces transverse momentum diffusion, the field response is encoded in the spatial off-diagonal components of the energy-momentum tensor. The corresponding metric components are of particular interest for us.

The quark-gluon plasma in the experimentally accessible regime behaves as a strongly coupled fluid. This is due to the fact that the energies in heavy-ion collisions are still too small to decrease the coupling constant significantly. Close to equilibrium, the QGP evolution and its reaction to perturbations long-scaled in space and time can be described by relativistic hydrodynamics. In contrast to an ideal fluid (IF) which is inviscid, fluctuations in the QGP are damped. In order to capture the dissipative behavior, the constituent equation for the ideal fluid energy-momentum tensor is replaced by a derivative expansion. At zeroth order there are the components of an ideal fluid, $T^\text{IF}_{\mu\nu}$, while the non-ideal fluid (NIF) components are contained in $T^\text{NIF}_{\mu\nu}$ \cite{BaierRSSS2008,LandauL1966}. Up to first order we obtain
\begin{eqnarray}
T_{\mu\nu}
&=& T^\text{IF}_{\mu\nu} + T^\text{NIF}_{\mu\nu}\nonumber\\
&=& \left(\epsilon \, u_\mu u_\nu + P\,\Delta_{\mu\nu}\right)
    +\left( -\eta\,\sigma_{\mu\nu}
    -\zeta\, \nabla_\rho u^\rho\, \Delta_{\mu\nu}
    +\mathcal{O}\!\left(\nabla^2\right)
\right).
\end{eqnarray}
Referring to a given reference frame, the fluid velocity is denoted by $u^\mu$, the energy density by $\epsilon$, and the pressure by $P$.
The projector $\Delta_{\mu\nu}$ and the shear tensor $\sigma_{\mu\nu}$ read
\begin{equation}
\Delta_{\mu\nu} \equiv g_{\mu\nu} + u_\mu u_\nu, \qquad
\sigma_{\mu\nu} \equiv 2 \nabla_{\langle \mu} u_{\nu \rangle}
\end{equation}
where angle brackets stand for the the symmetric traceless part of a tensor with respect to the corresponding indices.
The shear viscosity $\eta$ and bulk viscosity $\zeta$ are examples of transport coefficients. The latter describes the frictional resistance to compression or expansion. The former determines the transverse momentum diffusion which manifests itself, \textit{e.g.}, as the velocity profile of a fluid when applying shear. 
Consider a 3-dimensional case with a local Cartesian rest frame and a fluid between two surfaces parallel to the $x$-$z$ plane. A force in the $x$-direction applied tangentially to one surface with an amount $F$ per area $A$ leads to a velocity gradient $\partial_{y} u_{x}$,
\begin{equation}
-T_{yx} = \frac{F}{A} = \eta\, \sigma_{yx} = \eta\, \partial_{y} u_{x}.
\end{equation}
Thus the shear viscosity corresponds to a non-vanishing off-diagonal transverse component of the energy-momentum tensor.

\section{Calculation procedure}
\label{sec:CalculationProcedure}
As we have seen in the preceding section, shear is related to the $xy$-component of the energy-momentum tensor $T^{\mu\nu}$. For field states close to equilibrium the shear viscosity $\eta$ can be obtained by a Green-Kubo relation from the associated retarded Green's function $G^{xy,xy}_\text{R}$. 

We are interested in the behavior of the quark-gluon plasma under shear in the far-from-equilibrium regime and thus outside of the hydrodynamic range of applicability. This is the sketch of our approach: In the same way as above, we focus on $T^{xy}$ and apply linear response theory. We introduce a linearized metric perturbation, $h_{xy}$, in order to obtain the linearized answer in the expectation value of the energy-momentum tensor, $\left\langle \delta T^{xy} \right\rangle$. Since in our setup the corresponding background quantities vanish, $g^{\text{(bg)}}_{xy} = 0,\, \left\langle T^{\text{(bg)}\,xy} \right\rangle = 0$, we have 
$g_{xy} = h_{xy}$ and
$\left\langle T^{xy} \right\rangle 
= \left\langle \delta T^{xy} \right\rangle$.

The field state far from equilibrium is achieved by choosing a rapidly changing background solution in the bulk. Therefore time-translation invariance is broken, but the background is still spatially homogeneous and isotropic. This allows us to perform a Fourier transformation of the metric perturbation with respect to the transverse coordinates $\vec{x}$ to get rid of the corresponding derivatives. The equation of motion of the metric fluctuation is solved numerically with the infalling boundary condition at the black brane horizon. The one-point function $\left\langle T^{xy} \right\rangle$ is extracted from the boundary behavior of the bulk field. Choosing a special profile for the metric fluctuation, a Dirac delta distribution, the one-point function in presence of this source is the retarded Green's function. This allows us to study the two-point function in time \cite{BalasubramanianBCKKMTV2013,BanerjeeIJMR2016,Ishii2016}.

In our investigation, we want to drive a homogeneous and isotropic holographic plasma far from equilibrium by a sudden temperature rise. 
Therefore, we choose a 4-dimensional asymptotically AdS Vaidya spacetime in the bulk (cf.~\textit{e.g.}~\cite{Abajo-ArrastiaAL2010,CaceresKPT2014}). The metric in Poincar\'{e} form reads
\begin{equation}
\dd s^2
= r^2 \left[ -\left(1 - \frac{M\!\left(v\right)}{r^3} + \frac{Q^2\!\left(v\right)}{2r^4}\right) \, \dd v^2 +\dd x^2 +\dd y^2 +\frac{2}{r^2}\, \dd r \, \dd v \right].
\end{equation}
where $v$ is the ingoing Eddington-Finkelstein coordinate which coincides with ordinary time $t$ on the boundary. The AdS radial coordinate is denoted by $r$, and the coordinates $x$ and $y$ denote Cartesian non-compact transverse coordinates. The Vaidya solution generalizes the Schwarzschild black brane to the case of a time-dependent mass $M\!\left(v\right)$. We further allow for a time-dependent electric charge $\Qv$. The apparent horizon of the black brane is located at $r_\text{h}\!\left(v\right)$. Here and in the following, all quantities are normalized to appropriate powers of the AdS radius $L$.

\begin{figure}[!ht]
  \includegraphics[width=0.4\linewidth]{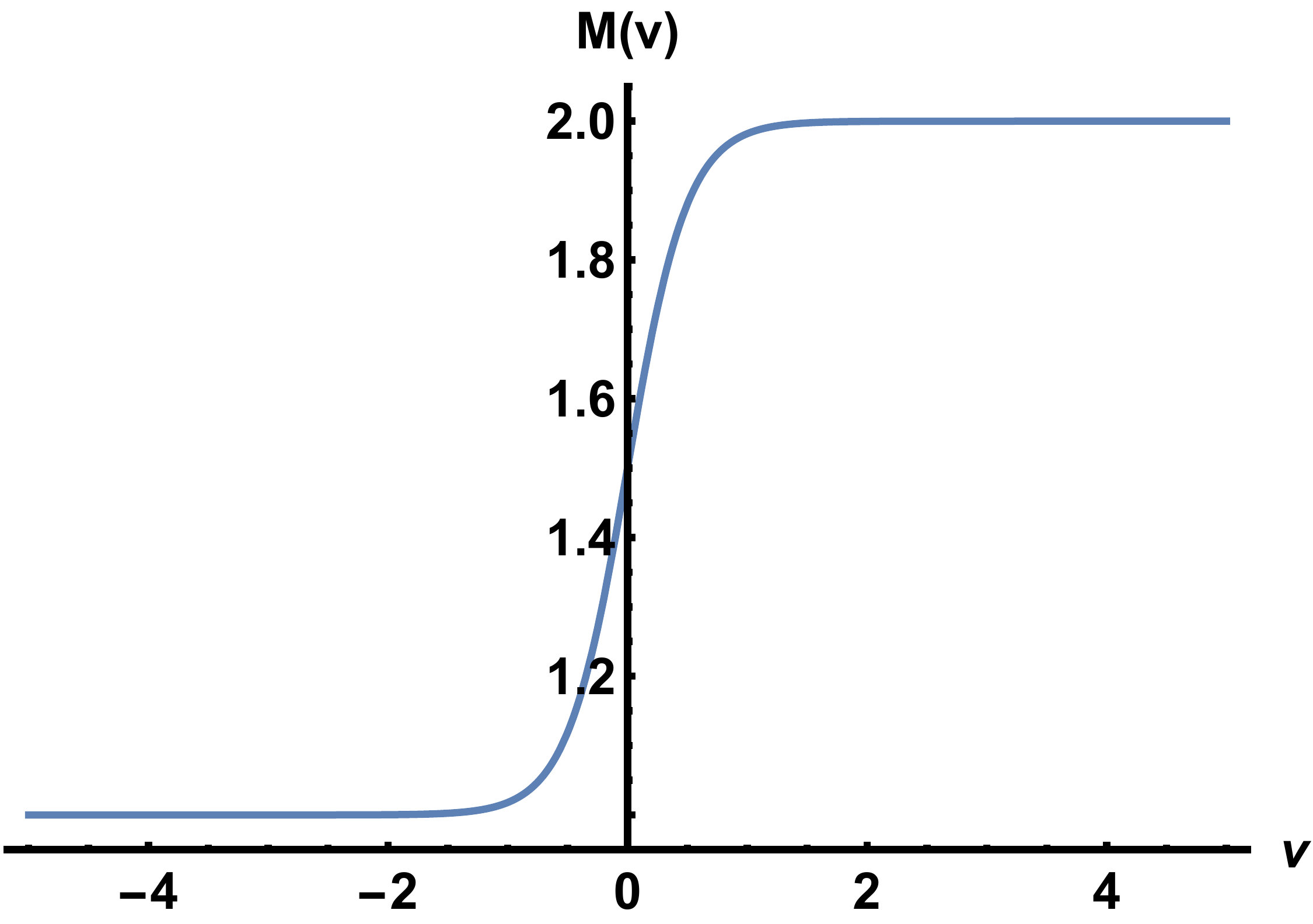}\hfill%
\begin{minipage}[b]{0.45\linewidth}%
\caption{\label{fig:ProfileMassShell}%
Sample mass profile $M\!\left(v\right)$ of the black brane due to the infall of null matter. It shows the smooth transition from the initial mass $m=1$ to the final mass $m+\mss=2$ centered around $\ts=0$ within a short time period determined by $\sigs=0.5$.}
\end{minipage}
\end{figure}

The time dependence is due to the infall of pressureless matter which propagates at the speed of light, known as null matter.
By choosing the mass of the brane to increase rapidly in a short null time interval, we provide the intended energy deposition. This is realized by the mass profile 
\begin{equation}
M\!\left(v\right)
	= m+\frac{1}{2}\, \mss \left(1 +\tanh\!\left(\frac{v-\ts}{\sigs}\right)\right)
\label{eq:MassProfile}
\end{equation}
where the steepness is parametrized by $\sigs$. The initial mass is denoted by $m$, the infalling mass by $\mss$. At time $\ts$ on the boundary, the temperature undergoes the phase of maximum increase rate. An example of this profile is shown in Fig.~\ref{fig:ProfileMassShell}.
The infalling null matter contributes to the bulk energy-momentum tensor and carries an electric current,
\begin{equation}
T_{mn}^{\text{bulk,\,null}}
= \frac{\mss\, \text{sech}^2\!\left( \frac{\ts-v}{\sigs} \right)\, r -2 \sigs \, \Qv \ddel{v} \Qv}{2 \sigs \, r^3}
  \,\delta_m^v\,\delta_n^v,
\qquad
J_\text{bulk,\,null}^m
= -\frac{\ddel{v} \Qv}{r^2}\,\delta_r^m.
\end{equation}

Furthermore, the black brane has a time-dependent electric charge $Q\!\left(v\right)$ and couples to the U($1$) gauge field $A_m$. The gauge field takes on a non-trivial value at the boundary which plays the role of the electric chemical potential $\mu\!\left(v\right)$ on the field theory side,
\begin{equation}
A_m\!\left(v,r\right)
= -\left(\mu\!\left(v\right)+\frac{Q\!\left(v\right)}{r}\right)\,\delta_m^v, \qquad
\mu\!\left(v\right) = -\frac{Q\!\left(v\right)}{r_\text{h}\!\left(v\right)}.
\end{equation}

\section{Metric fluctuation}
\label{sec:MetricFluctuation}
To study the properties of the plasma, we introduce linearized perturbations of the background fields, namely $h_{\mu\nu}$ associated to the metric and $a_\mu$ associated to the gauge field. We apply the radial gauge in which the $r$-components of the fluctuations vanish. In the linearized equations of motion, $h_{xy}$ decouples from the other fluctuations. Using the inverse radial coordinate $z \equiv 1/r$ and rescaled metric fluctuation $\hxyh \equiv \hxy/z^2$, the equation of motion reads
\begin{eqnarray}
	0&=&-2z \left(1 -M\!\left(v\right)\,z^3 +\frac{\Qsqv}{2}\,z^4 \right) \ddel{z}^2\hxyh  \nonumber\\
	&&{}+\left(4 +2M\!\left(v\right)\,z^3 -2 \Qsqv\,z^4\right) \ddel{z}\hxyh -4 \,\ddel{v}\hxyh +4z \,\ddel{v}\ddel{z}\hxyh.
	\label{eq:dd_Delta_Minus}
\end{eqnarray}

\begin{figure}[!ht]
  \includegraphics[width=0.45\linewidth]{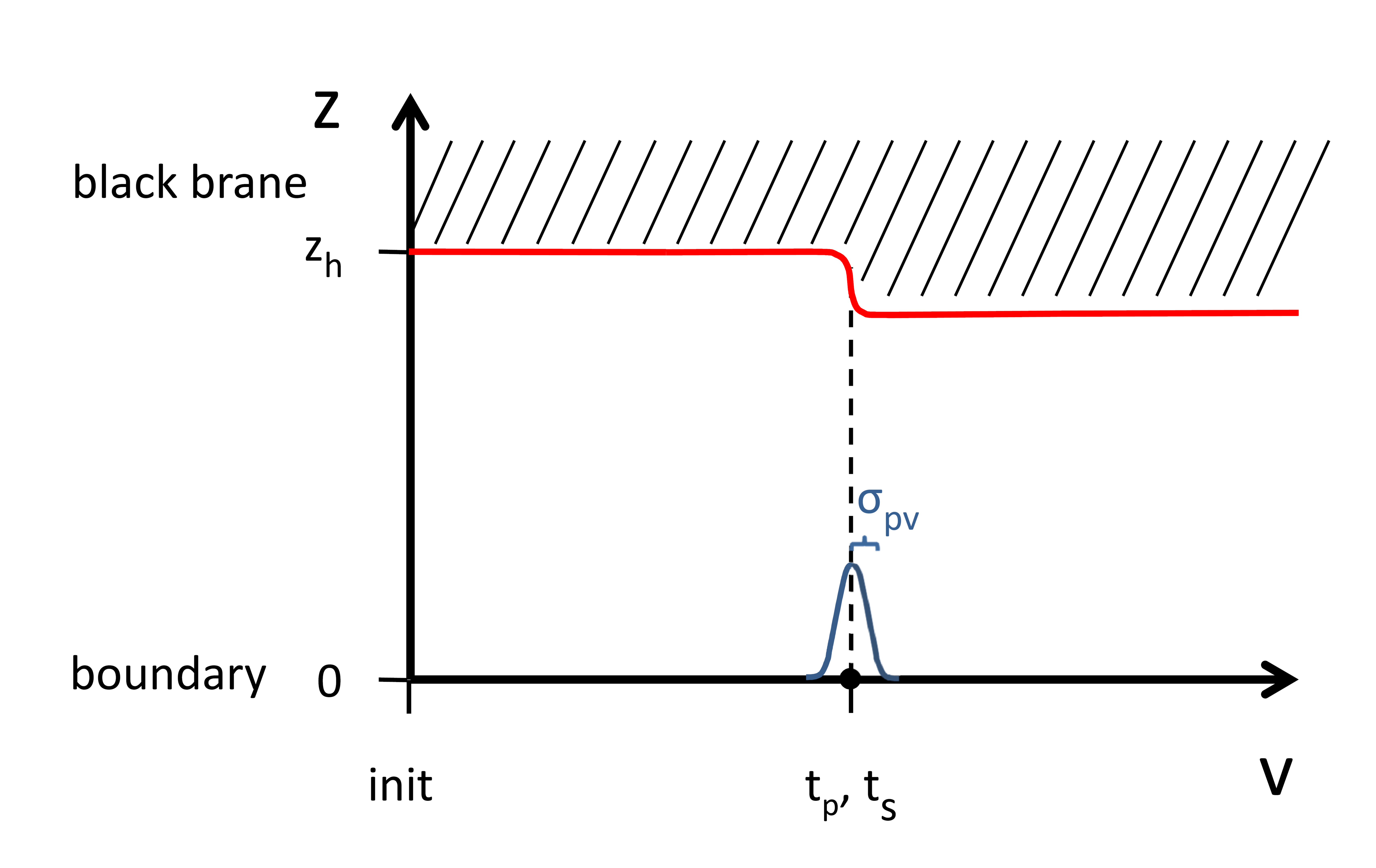}\hfill%
\begin{minipage}[b]{0.45\linewidth}%
\caption{\label{fig:Skizze_Setup_nachher}%
Sketch of the spacetime. At boundary time $\ts$ the mass infall reaches its maximum, while $\tp$ defines the time interval of width $\sigpv$ at which the metric fluctuation is sourced. The black brane horizon is given by $z_\text{h}$ (red line).}
\end{minipage}
\end{figure}

The equation of motion can be solved numerically. The analytic boundary expansion of the field in terms of the radial coordinate reveals two independent coefficients, the source $h_{xy}^{(0)}$ of the boundary operator $T^{xy}$, as well as, at subleading order, the expectation value $\left\langle T^{xy} \right\rangle$ up to contact terms. According to linear response theory, we can obtain the Green's function in position space by evaluating the one-point function in presence of a Dirac delta-like source $\hat{h}^\text{pulse}_{xy}$,
\begin{eqnarray}
\left\langle T^{xy}\!\left(t_1,\,\vec{x}_1\right) \right\rangle_{h}
&=& -\int\!\dd^{d} \xi\; G_\text{R}^{xy,xy}\!\left(t_1,\,\tau;\,\vec{x}_1-\vec{\xi}\right) \, \underbrace{\hat{h}^\text{pulse}_{xy}\!\left(\xi\right)}_{= \delta\left(\tau-t_2\right)\,\delta\left(\vec{\xi}-\vec{x}_2\right)} \nonumber\\[-2ex]
&=& - G_\text{R}^{xy,xy}\!\left(t_1,\,t_2;\,\vec{x}_1-\vec{x}_2\right).
\end{eqnarray}
A sketch of the spacetime including the metric fluctuation is presented in Fig.~\ref{fig:Skizze_Setup_nachher}. In order to implement the boundary source in the numerical solution procedure, we approximate it by a Gaussian profile,
\begin{equation}
\hat{h}^\text{pulse}_{xy}\!\left(v\right)
= \Ap\,\delta\!\left(v-\tp\right)
\simeq \frac{\Ap}{\sqrt{2\pi}\,\sigpv}\, \exp\!\left( -\frac{1}{2} {\left(\frac{v-\tp}{\sigpv}\right)}^2 \right).
\end{equation}
Here $\Ap$ denotes the integrated amplitude of the pulse with width $\sigpv$ at boundary time $\tp$. In this way, we can deduce the retarded Green's function from the boundary behavior of the fluctuation in an easy way. Currently, the results of this numerical calculation are being checked and analyzed.

\section{Conclusion}
\label{sec:Conclusion}
The AdS/CFT correspondence allows the characterization of a plasma far from equilibrium. The position-space Green's function can be a measure of the shear property of the time-dependent plasma. We investigated a 4-dimensional charged asymptotically AdS Vaidya spacetime as a gravitational dual for a homogeneous and isotropic excitation of the plasma. We derived the Green's function in terms of the one-point function in the presence of a Dirac delta-shaped source. Our investigation paves the way for a deeper understanding of the quark-gluon plasma far from equilibrium via the AdS/CFT correspondence.

\ack
MFW wants to thank the organizers of the \textit{3rd Karl Schwarzschild Meeting on Gravitational Physics and the Gauge/Gravity Correspondence} for an highly interesting and stimulating conference. He is grateful to the Stiftung Polytechnische Gesellschaft Frankfurt am Main. MK was supported in part by the UA ``Research in Elementary Particle Physics'' DOE grant. The work of PN has been supported by the project ``Evaporation of the microscopic black holes'' of the German Research Foundation (DFG) under the grant NI 1282/2-2.

%\appendix
%\setcounter{section}{1}

%\section*{References}

\providecommand{\newblock}{}

\end{document}